\documentclass[11pt]{article}
\usepackage{latexsym}
\usepackage{amssymb}
\usepackage{amsmath}
\usepackage[hypertex]{hyperref}
\usepackage{graphicx}% Include figure files

\begin{document}

\begin{flushright}
SLAC-PUB-11897\\
hep-ph/0606129 \\
\end{flushright}

\vspace{2.5cm}

\begin{center}

{\bf\LARGE Dynamical GUT breaking and 
{\boldmath $\mu$}-term driven supersymmetry breaking} \\

\vspace*{1.5cm}
{\large Ryuichiro Kitano} \\
\vspace*{0.5cm}

{\it Stanford Linear Accelerator Center, Stanford University,
                Stanford, CA 94309} \\

\vspace*{0.5cm}

\end{center}

\vspace*{1.0cm}

\begin{abstract}
Models for dynamical breaking of supersymmetric grand unified theories
are presented. The doublet-triplet splitting problem is absent since the
Higgs doublet superfields can be identified with the massless mesons of
the strong gauge group whereas there are no massless states
corresponding to the colored Higgs fields.  Various strong gauge groups
SU($N_c$), Sp($N_c$) and SO($N_c$) are examined. In a model with SO(9)
strong gauge group, adding $\mu$-term for the Higgs fields triggers to
break supersymmetry in a meta-stable vacuum.  The pattern of the
supersymmetry breaking parameters is predicted to be of the
gauge-mediation type with modifications in the Higgs sector.

\end{abstract} 

%%%%%%%%%%%%%%%%%%%%%%%%%%%%%%%%%%%%%%%%%%%%%%%%%%%%%%%%%%%%%%%%%%%%%%%%%%%%
\newpage
\baselineskip 18pt

\section{Introduction}

The Higgs boson, which has not been observed yet, is the most mysterious
particle in the standard model although it plays an important role:
electroweak symmetry breaking and the origin of the fermion masses.
Successful electroweak symmetry breaking needs a negative mass squared
for this particle and its size must be anomalously small compared to the
cut-off scale of the theory.
This situation motivated us to consider the supersymmetric standard
model to protect the mass parameter from large quantum correction, but
the Higgs particle is still left mysterious.
The Higgs boson (or Higgsino) mass parameter, the $\mu$-term, cannot be
protected by gauge symmetry or supersymmetry (SUSY) although quantum
corrections are successfully removed.

Embedding the supersymmetric standard model into grand unified theories
(GUT), motivated by gauge coupling unification~\cite{Dimopoulos:1981zb},
makes the Higgs particle more mysterious. Even for the smallest group
for grand unification, SU(5)~\cite{Georgi:1974sy}, the Higgs fields do
not fit into a complete multiplet of the symmetry group, and therefore
we need extra particles to appear at the GUT scale $M_G \sim
10^{16}$~GeV. One mysterious feature is that, it is not easy to realize
this situation with nearly massless Higgs fields in models with the GUT
symmetry breaking by the Higgs mechanism. The splitting of masses
between the Higgs fields and extra particles must be done by a coupling
to the fields whose vacuum expectation values (VEV) break the GUT
symmetry group, but in simple models it gives masses to the Higgs fields
of the order of $M_G$. This is the famous doublet-triplet splitting
problem.
Another mysterious feature of SUSY-GUT models is the absence or
suppression of dimension-five proton-decay operators. For example, in
the simplest SU(5) model, the Higgs particles are embedded into the
${\bf 5}$ and ${\bf \bar{5}}$ representations which contain two colored
partner of the Higgs fields $H_C$ and $\bar{H}_C$. If these colored
Higgs fields have masses by pairing to each other, i.e., $W \ni m_C H_C
\bar{H}_C$, dimension-five operators suppressed by the scale $m_C$ are
generated by integrating out $H_C$ and
$\bar{H}_C$~\cite{Weinberg:1981wj}. It has been studied that the colored
Higgs mass $m_C$ has to be quite large $m_C \gtrsim 10^{17}$~GeV, which
is disfavored by the unification of the gauge
couplings~\cite{Goto:1998qg}.

The above two problems, the doublet-triplet splitting and the proton
decay, are actually related and there is a simple solution to these
problems.
If we are to avoid dimension-five proton-decay operators such as $QQQL$
in the superpotential, with $Q$ and $L$ being the quark and lepton
doublets, an easy way is to impose a symmetry under which matter fields
are charged, e.g., both $Q$ and $L$ have charge unity.
In SU(5) SUSY GUTs, this means that both of the Higgs fields in ${\bf
5}$ and ${\bf \bar{5}}$ representations have charge $-2$ in order to
have Yukawa interactions, and thus the mass term is forbidden.
Now, to give mass terms to the colored Higgs fields while keeping the
symmetry unbroken we need to introduce another pair of ${\bf 5}$ and
${\bf \bar{5}}$ field which have charge $+2$, but in this case, we have
either zero or four Higgs-doublet fields at low energy which is
unacceptable. Of course, adding another pair of ${\bf 5}$ and ${\bf
\bar{5}}$ field with charge $-2$ results in extra massless colored Higgs
fields. Therefore, in order to have only two Higgs doublets while
forbidding the proton decay by continuous symmetry, we need to repeat
the procedure of adding ${\bf 5}$ and ${\bf \bar{5}}$ forever and end up
with an infinite number of Higgs fields in ${\bf 5}$ and ${\bf \bar{5}}$
representations~\cite{watari}.

While an infinite number of particles sounds unreasonable in field
theory, it is quite possible to realize this situation in models with
extra-dimensions. The infinite number of particles is identified with
the Kaluza-Klein tower of the fields which are propagating into the bulk
of the extra-dimension. Indeed, simple GUT models have been constructed
in higher dimensional space-time, where the boundary condition breaks
the GUT symmetry and there is no doublet-triplet
splitting~\cite{Kawamura:2000ev} or the proton decay
problem~\cite{Altarelli:2001qj}. In this picture, the Higgs particles
become less mysterious. They are just bulk fields.

On the other hand, there is another familiar mechanism of having an
infinite tower of particles in field theory, that is actually happening
in QCD. When an asymptotically free gauge theory becomes strong at low
energy, the effective theory below that scale is described by
gauge-singlet particles such as mesons and baryons. These particles have
also an infinite tower of excitation states.
This fact naturally leads us to think of the possibility of realizing
Higgs particles as composite fields in some strongly coupled gauge
theory which breaks the GUT symmetry dynamically.
This question is also interesting from the viewpoint of the AdS/CFT
correspondence~\cite{Maldacena:1997re, Arkani-Hamed:2000ds}. The
extra-dimensional GUT models above may be interpreted as a dual picture
of the strongly coupled theory. The explicit gauge symmetry breaking in
the extra-dimensional picture may be justified by the presence of viable
dynamical GUT breaking models.

Constructing GUT models associated with a strongly coupled gauge theory
have been attempted by the group of Hotta, Izawa and
Yanagida~\cite{Yanagida:1994vq,Hotta:1995ih,Hotta:1996qb,Izawa:1997br}.
(See also~\cite{Cheng:1997fk} for subsequent works.)
Various
gauge groups for the strong interaction, SU(3) ($\times$
U(1))~\cite{Yanagida:1994vq,Hotta:1995ih}, SU(5)~\cite{Izawa:1997br} and
SO(6)~\cite{Hotta:1996qb}, were considered.
It was found that the doublet-triplet splitting can be easily realized
via the missing partner mechanism while preserving an (anomalous) U(1)
symmetry which forbids dimension-five proton decays.
Along a similar line, a model with Sp(2) gauge group has also been
constructed recently in Ref.~\cite{Kitano:2005ez} where the model is
quite simplified. (Our convention is such that Sp(1) $\simeq$ SU(2).)
The model consists of six flavors of quarks of Sp(2) and the five of
flavors are identified with the ${\bf 5}$ and ${\bf \bar{5}}$
representation fields of the SU(5) GUT. The other flavor turns out to be
a (constituent of) the Higgs doublets in low energy.
In this model, the conformal field theory (CFT) nature of the Sp(2)
interaction above the GUT scale plays a crucial role.
A similar approach in warped extra-dimension can also be found in
Ref.~\cite{Nomura:2006pn}.

In this paper, we consider a generalization of the Sp(2) model.  We find
that models with Sp($N_c$) with $N_c=2$, and SO($N_c$) with $6 \leq N_c
\leq 9$ work for dynamical GUT breaking while having massless doublet
Higgs fields, and no viable SU($N_c$) group is found under the
assumption on the particle content and superpotential.
Of particular interest is the case with SO($N_c$) gauge group. There is
no exotic particle left massless without adding superpotential terms to
remove those particles.

With the success of the doublet-triplet splitting while forbidding the
proton decay, the final missing piece for the Higgs mystery is the
finite $\mu$-term.
In the gravity mediated supersymmetry breaking
scenario~\cite{Chamseddine:1982jx}, it is possible to obtain a correct
size of the $\mu$-term in a simple way~\cite{Giudice:1988yz}. However,
we find an alternative interesting possibility in the SO(9) model.
Instead of solving the $\mu$-problem, if we add a small $\mu$-term in
the superpotential, supersymmetry breaks down at the intermediate scale
$F \sim \mu M_G$.
Although this is not the true vacuum, it is shown to be
meta-stable~\cite{Intriligator:2006dd}.
It is amusing that the $\mu$-term can drive supersymmetry breaking,
which is the opposite direction to the usual thought.
The smallness of the $\mu$-term is ``explained'' by demanding a low
supersymmetry breaking scale.
The Higgs fields can be responsible not only for electroweak symmetry
breaking but also for GUT and supersymmetry breaking.

There is an interesting possibility for the nature of the small
$\mu$-term.  The small $\mu$-term added by hand can come from a negative
cosmological constant term in the supergravity action. By the
Giudice-Masiero mechanism~\cite{Giudice:1988yz}, the (supersymmetric)
cosmological constant term induces a $\mu$-term in the presence of a
particular K{\" a}hler potential term. This $\mu$-term, in turn, drives
supersymmetry breaking which gives positive contribution to the vacuum
energy and cancels the net cosmological constant.

We start the discussion of dynamical GUT breaking in Section 2, there
the general set-up is defined. We analyze a successful model, the SO(9)
model, in Section 3. In Section 4, the mechanism of $\mu$-term driven
supersymmetry breaking is presented and we discuss the mediation of the
supersymmetry breaking to our sector. The generation of the $\mu$-term
through the cosmological constant is discussed in Section 5.

\section{Dynamical GUT breaking}

In Ref.~\cite{Kitano:2005ez}, a simple model for the dynamical GUT
breaking was constructed based on an Sp(2) gauge theory.
We study a generalization of the model with various gauge groups:
SU($N_c$), Sp($N_c$) and SO($N_c$).
Models with Sp($N_c$) and SO($N_c$) with a certain range of $N_c$ is
found to be viable but no viable SU($N_c$) group is found.
Although they are not successful models, we start with the discussion of
the SU($N_c$) models in which we can see the essential features of this
class of models.

\subsection{SU($N_c$) models}

\begin{table}
\begin{center}
\begin{tabular}{ccc}
 & SU($N_c$) & SU(5)$_{\rm GUT}$  \\ \hline \hline
 $Q$ & { $N_c$} & {\bf 5}  \\
 ${\bar{Q}}$ & { $\bar{N}_c$} & ${\bf \bar{5}}$  \\
 $T$ & { $N_c$} & {\bf 1}  \\
 $\bar{T}$ & { $\bar{N}_c$} & {\bf 1}  \\ \hline
\hline
\end{tabular}
\end{center}
\caption{
The particle content of the SU($N_c$) model. 
}
\label{tab:content}
\end{table}

The model consists of six flavors, and five of which carry SU(5)$_{\rm
GUT}$ quantum numbers as listed in Table~{\ref{tab:content}}. The quarks
and leptons in the standard model are not charged under SU($N_c$) and
are unified usually as ${\bf 10}$ and ${\bf \bar{5}}$ of SU(5)$_{\rm
GUT}$.
We introduce a superpotential for $Q$ and $\bar{Q}$:
\begin{eqnarray}
 W = m {\rm Tr} (Q \bar{Q})
 - \frac{1}{M}{\rm Tr} [(Q \bar{Q}) (Q \bar{Q})]
 + \cdots \ ,
\label{eq:superpotential}
\end{eqnarray}
where $(Q \bar{Q})$ is the SU($N_c$) singlet $5 \times 5$ matrix, and
`$\cdots$' represents other higher dimensional operators such as $({\rm
Tr} (Q \bar{Q}))^2$ and those are not important for the discussion.
It is essential for the masslessness of the Higgs fields that the
superfields $T$ and $\bar{T}$ do not have a superpotential at tree level
since the Higgs fields will be identified with the meson fields $H \sim
Q \bar{T}$ and $\bar{H} \sim \bar{Q} T$.

Before the analysis at the quantum level, it is helpful for the
understanding of the model to discuss what happens at the classical
level.
The classical analysis is valid for $\Lambda \ll M_G$ with $\Lambda$
being the dynamical scale of SU($N_c$). In this case, the picture
becomes similar to models with product group
unification~\cite{Barbieri:1994jq}.
At the classical level, there are vacua with rank$(Q \bar{Q}) = 0$ to
$\min [5,N_c]$ which satisfy the conditions of $F_Q = F_{\bar{Q}} = 0$.
We are interested in the vacuum with rank$(Q \bar{Q}) = 2$:
\begin{eqnarray}
 (Q \bar{Q}) = \left(
\begin{array}{ccccc}
 0& & & & \\
 &0 & & & \\
 & &0 & & \\
 & & &v^2 & \\
 & & & &v^2 \\
\end{array}
\right)\ ,
\label{eq:vacuum}
\end{eqnarray}
where $v^2 = m M / 2 $. At the vacuum, the SU($N_c$) $\times$
SU(5)$_{\rm GUT}$ gauge symmetry is broken down to SU($N_c - 2$)
$\times$ SU(3)$_C$ $\times$ SU(2)$_L$ $\times$ U(1)$_Y$ for $N_c \geq 3$
and the electroweak SU(2)$_L$ $\times$ U(1)$_Y$ is the diagonal subgroup
of those in SU($N_c$) and SU(5)$_{\rm GUT}$.
Note that the vanishing components are not a consequence of the
fine-tuning. The corresponding components in $Q$ and $\bar{Q}$ are
charged under the unbroken gauge symmetry and that ensures the absence
of linear terms in the potential, i.e., stable (or flat directions).
Since the low energy SU(2)$_L$ $\times$ U(1)$_Y$ partly comes from
SU($N_c$), two of the components in $T$ and $\bar{T}$ transforms in
exactly the same way as the Higgs fields in low energy whereas there is
no colored component in $T$ or $\bar{T}$. Therefore the double-triplet
splitting problem and proton decay mediated by the colored Higgs are
absent. The rest of the components in $T$ and $\bar{T}$ are fundamental
and anti-fundamental representations of SU($N_c-2$) and charged under
U(1)$_Y$.
All the components in $Q$ and $\bar{Q}$ are either eaten by gauge fields
of the broken symmetry or obtain masses from the superpotential.
The fate of the exotic particles in $T$ and $\bar{T}$ depends on the
dynamics of the unbroken SU($N_c-2$) group below the dynamical scale.
Of course, with $\Lambda \ll v$, this is not a ``unified'' model. The
three gauge coupling constants do not meet at the GUT scale since the
embeddings of SU(3)$_C$ and SU(2)$_L$ $\times$ U(1)$_Y$ are different.
The real unification picture arises when $\Lambda \gtrsim v$, where
quantum effects are important.

At the quantum level, the low energy physics is not very different, but
some of the vacua are lifted.
In particular, it is interesting to note that dynamical symmetry
breaking has to happen once we take into account quantum
effects~\cite{Hotta:1996qb}.  With the above superpotential, the low
energy theory of rank$(Q \bar{Q}) = 0$ vacuum is SU($N_c$) with one
flavor $T$ and $\bar{T}$, which does not have the ground
state~\cite{Seiberg:1994bz}.
The stability of the classical vacua of rank$(Q \bar{Q}) = 2$ depends on
$N_c$. We show below that there is no $N_c$ which is viable for low
energy phenomenology.

For $N_c = 2$, the theory is not asymptotically free and the classical
analysis is valid in low energy. However, in this case, the vacuum in
Eq.~(\ref{eq:vacuum}) breaks the gauge symmetry into SU(3) $\times$
SU(2) which is not acceptable.

For $N_c = 3$, the quantum effect is easier to analyze in the dual gauge
theory~\cite{Seiberg:1994pq}. It is again an SU(3) gauge theory but with
a superpotential:
\begin{eqnarray}
 W &=& m {\rm Tr} M_{Q \bar{Q}} 
- \frac{1}{M} {\rm Tr} (M_{Q \bar{Q}} M_{Q \bar{Q}})
+ \cdots
\nonumber \\
&+& \frac{1}{\hat{\Lambda}} \bar{q} M_{Q \bar{Q}} {q} 
+ \frac{1}{\hat{\Lambda}} {H} \bar{q} {t} 
+ \frac{1}{\hat{\Lambda}} \bar{H} {q} \bar{t} 
+ \frac{1}{\hat{\Lambda}} {S} {t} \bar{t} \ ,
\end{eqnarray}
where mesons are identified with the quark bilinears in the original
(electric) theory: $M_{Q \bar{Q}} \sim Q \bar{Q}$, $H \sim Q \bar{T}$,
$\bar{H} \sim \bar{Q} T$ and $S \sim T \bar{T}$. These mesons are
singlets under SU(3) and transforms as ${\bf 1} + {\bf 24}$, ${\bf 5}$,
${\bf \bar{5}}$ and ${\bf 1}$ under SU(5)$_{\rm GUT}$,
respectively. Dual quarks $q$, $t$ and anti-quarks $\bar{q}$, $\bar{t}$
transforms as $q: (\overline{\bf 3}, {\bf 5})$, $t: (\overline{\bf 3},
{\bf 1})$, $\bar{q}: ({\bf 3}, {\bf \bar{5}})$ and $\bar{t}: ({\bf 3},
{\bf {1}})$ under SU(3) $\times$ SU(5)$_{\rm GUT}$. The parameter
$\hat{\Lambda}$ has dimension one since mesons have dimension two.
Amazingly, this is almost identical to the model proposed in
Ref.~\cite{Hotta:1995ih}.
In the vacuum of our interest:
\begin{eqnarray}
 M_{Q \bar{Q}} = \left(
\begin{array}{ccccc}
 0& & & & \\
 &0 & & & \\
 & &0 & & \\
 & & &v^2 & \\
 & & & &v^2 \\
\end{array}
\right)\ ,
\label{eq:vacuum-2}
\end{eqnarray}
SU(5)$_{\rm GUT}$ is broken down to the standard model gauge group, and
two of the dual quarks in $q$ and $\bar{q}$ obtain masses. After
integrating out the massive quarks, the theory becomes an SU(3) gauge
theory with four flavors, which is a confining theory with a
superpotential~\cite{Seiberg:1994bz}:
\begin{eqnarray}
 W &=& m {\rm Tr} M_{Q \bar{Q}} 
- \frac{1}{M} {\rm Tr} (M_{Q \bar{Q}} M_{Q \bar{Q}})
+ \cdots
\nonumber \\
&+& 
\frac{1}{\hat{\Lambda}} M_{Q \bar{Q}}^{(\rm 3 \times 3)} M_{q \bar{q}}^{(\rm 3 \times 3)}
+ \frac{1}{\hat{\Lambda}} {H_C} \bar{H}^\prime_C
+ \frac{1}{\hat{\Lambda}} \bar{H}_C {H}^\prime_C 
+ \frac{1}{\hat{\Lambda}} {S} S^\prime \ ,
\nonumber \\
&-& 
\frac{1}{v^2 \hat{\Lambda}} H_D \bar{H}_D S^\prime + \cdots
\nonumber \\
&+& 
\frac{{\rm det} M^{(4 \times 4)}}{\widetilde{\Lambda}^5} 
+ \frac{1}{\widetilde{\Lambda}^{5}} B_{(4)} M^{(4 \times 4)}
\bar{B}_{(4)}.
\label{eq:dual-super}
\end{eqnarray}
The superscript $(3 \times 3)$ represents the $3 \times 3$ meson matrix
made of colored parts of $Q \bar{Q}$ or $q \bar{q}$.
The fields $H^\prime$, $\bar{H}^\prime$ and $S^\prime$ are mesons made
of dual quarks; $H^\prime \sim q \bar{t}$, $H^\prime \sim \bar{q} t$ and
$S^\prime \sim t \bar{t}$.  The subscripts $C$ and $D$ represent the
SU(3)$_C$ colored and the SU(2)$_L$ doublet part of the corresponding
meson fields, respectively. $M^{(4 \times 4)}$ is the matrix:
\begin{eqnarray}
 M^{(4 \times 4)} = \left(
\begin{array}{c|c}
 M_{q \bar{q}}^{(3 \times 3)} & H_C^\prime \\ \hline
 \bar{H}_C^{\prime T} & S^\prime\\
\end{array}
\right),
\end{eqnarray}
and the baryons are 
\begin{eqnarray}
 B_{(4)} = \left(
\begin{array}{c}
 B_C \\
 B^- \\
\end{array}
\right),\ \ \ 
 \bar{B}_{(4)} = \left(
\begin{array}{c}
 \bar{B}_C \\
 B^+ \\
\end{array}
\right),
\end{eqnarray}
with the identification of $B_C \sim q q t$, $B^- \sim q q q$,
$\bar{B}_C \sim \bar{q} \bar{q} \bar{t}$ and $B^+ \sim \bar{q} \bar{q}
\bar{q}$. Finally $\widetilde{\Lambda}$ is the dynamical scale of the
four-flavor SU(3) theory.

\begin{figure}[t]
 \begin{center}
  \includegraphics[height=6.5cm]{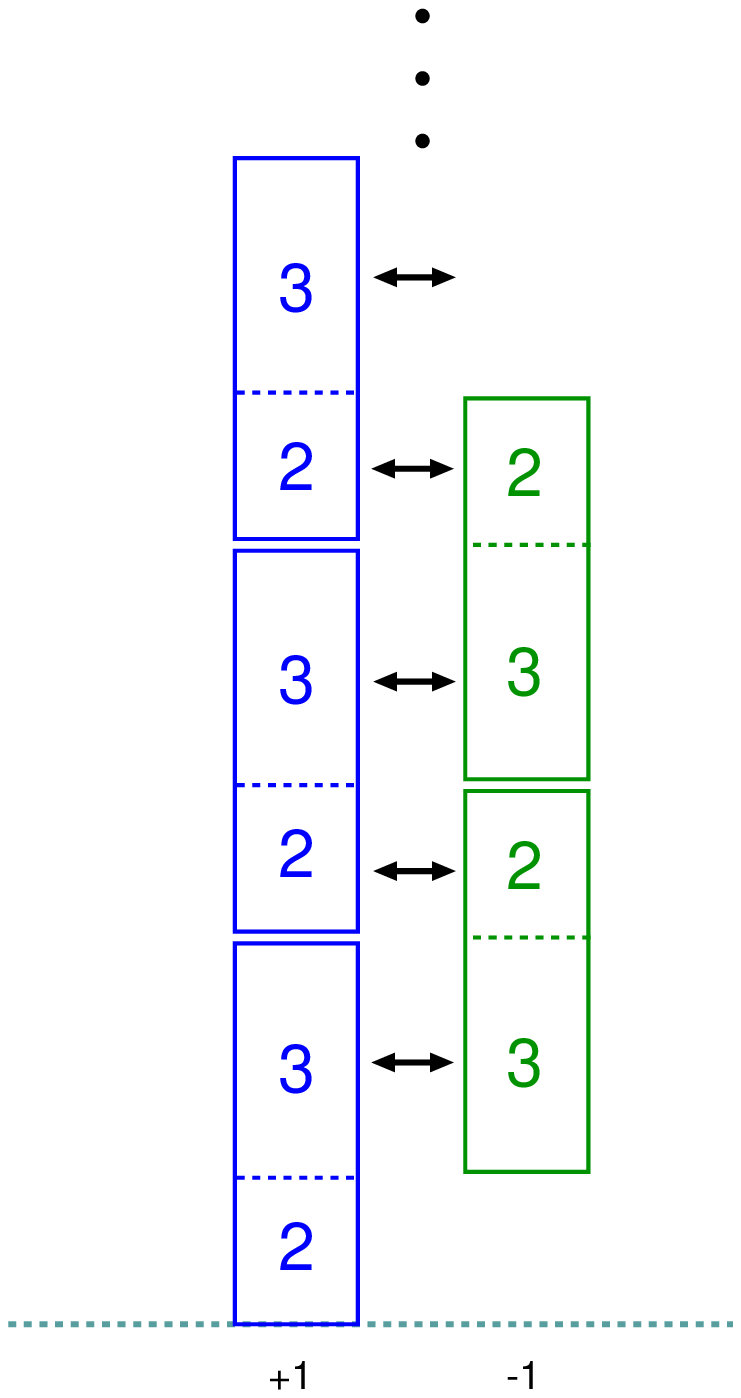}
\hspace*{3mm}
  \includegraphics[height=6.5cm]{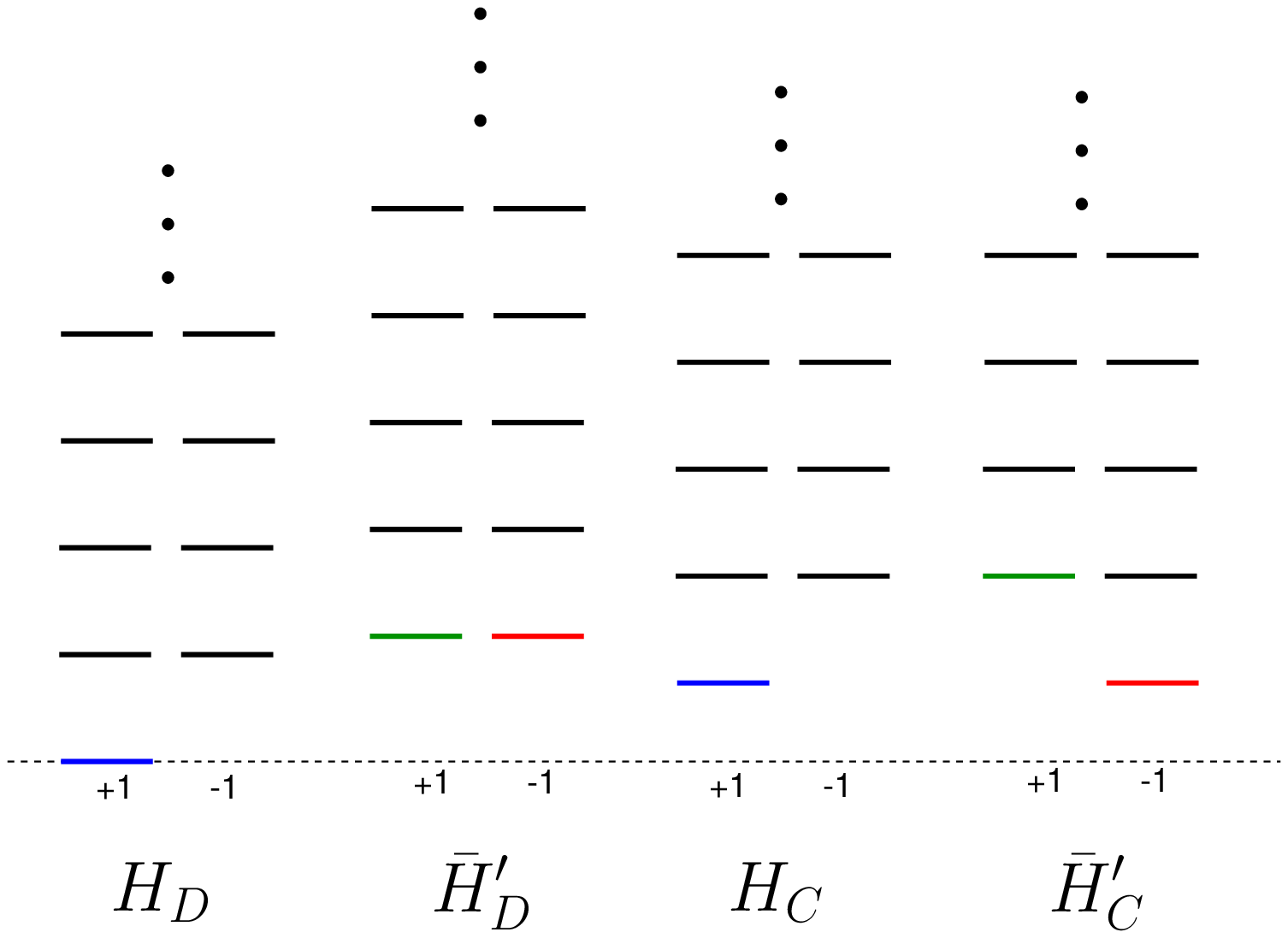}
\end{center}
\caption{A schematic view of the hadron spectrum of the model. Only
 doublet part $H_D$ remains massless whereas the colored Higgs $H_C$
 have masses by pairing up with $\bar{H}_C^\prime$ }
 \label{fig:spectrum}
\end{figure}

Note here that the doublet Higgses $H_D$ and $\bar{H}_D$ do not obtain a
mass term as long as $S^\prime = 0$ whereas the colored Higgses $H_C$
and $\bar{H}_C$ already have mass terms accompanied with the dual mesons
$H_C^\prime$ and $\bar{H}_C^\prime$.
Indeed $S^\prime = 0$ is ensured by the condition of $F_S = 0$. The
missing partner for the doublets, say $H_D^\prime$ and
$\bar{H}_D^\prime$, have masses since the dual quarks are massive. 
This is the realization of the doublet-triplet splitting with an
infinite number of particles.
If we define the $T$-number with the charge assignment $Q:0$,
$\bar{Q}:0$, $T:+1$ and $\bar{T}:+1$, the Higgs fields $H$ and $\bar{H}$
have the same $T$-number, $+1$.
In this case, an infinite tower of the Higgs fields should be necessary
as in Fig.~\ref{fig:spectrum} (left).
We need an extra Higgs field with charge $-1$ to make the colored Higgs
massive, but it introduces an additional unwanted massless doublet.
Repeating this procedure forever is the only possibility of realizing
doublet-triplet splitting in this situation.
These infinite particles are realized here by the hadron tower of the
Higgs fields. The mismatching of the level, i.e., no zero mode only for
doublet part of $H^\prime$ and $\bar{H}^\prime$, happened because of the
mass of the constituent quarks $q_D$ and $\bar{q}_D$ without violating
$T$-number. A schematic picture of the hadron spectrum is shown in
Fig.~\ref{fig:spectrum} (right).

In fact, in this SU(3) model, the situation is a little bit different
from the story in the Introduction.
If the $T$-number violating term is absent in the superpotential, the
dangerous dimension-five proton-decay operators are forbidden.
However, since $T$-number is anomalous with respect to the SU(3) gauge
theory, the non-perturbatively generated superpotential in
Eq.~(\ref{eq:dual-super}) violates $T$-number. As we see later, $M_{q
\bar{q}}^{(3 \times 3)}$ acquires a VEV and it gives $T$-number
violating masses to the colored Higgs fields.
Consequently, dimension-five proton decay operators are generated by the
colored-Higgs-exchange diagrams as usual.

The stability of the vacuum can be checked by solving the $F=0$
conditions for all the fields. We can find a solution with
Eq.~(\ref{eq:vacuum-2}) and
\begin{eqnarray}
 M_{q \bar{q}}^{(3 \times 3)} = \left(
\begin{array}{ccc}
- m \hat{\Lambda} & & \\
 &- m \hat{\Lambda} & \\
 & &- m \hat{\Lambda} \\
\end{array}
\right),
\label{eq:mqq-dual}
\end{eqnarray}
\begin{eqnarray}
 \frac{S}{\hat{\Lambda}} 
+ \frac{B^+ B^-}{\widetilde{\Lambda}^{5}}
= - \frac{m^3 \hat{\Lambda}^3 }{\widetilde{\Lambda}^5 }\ .
\end{eqnarray}
The second equation indicates that the vacuum is not uniquely determined
and there is a flat direction. Correspondingly, there are massless
particles $B^+$ and $B^-$ which are charged under U(1)$_Y$.
This is the same situation as in the classical analysis, in which the
remaining gauge symmetry is just the standard model gauge group but a
pair of U(1)$_Y$ charged particles originate from $T$ and $\bar{T}$ are
left massless in addition to the Higgs doublets. In order to avoid the
massless charged baryons, we need to add a mass term for $T$ and
$\bar{T}$, but that also makes the Higgs doublets heavy. Therefore, the
$N_c = 3$ case is not acceptable.

For $N_c = 4$, there is a hope that the exotic states confine and form
standard model singlet states so that it is phenomenologically
viable. However, unfortunately, it is not the case. As in the case of
$N_c= 3$, we can analyze the model by taking a dual gauge group and
integrate out the heavy flavors. For $N_c = 4$, the dual theory becomes
SU(2) with four flavors. As expected, doublet-triplet splitting happens
in the same way as above.
By taking the dual again and going back to the electric theory, we find
another SU(2) theory with four flavors and superpotential interactions.
The solution of the $F=0$ equations can be found with the same vacuum in
Eq.~(\ref{eq:vacuum-2}) and (\ref{eq:mqq-dual}), which gives mass terms
for all the quarks except for $T$ and $\bar{T}$. The low energy theory
in this vacuum is then an SU(2) gauge theory with one flavor which has
no ground state. Although we could avoid the charged exotic state, the
vacuum is lifted at the quantum level.

For $N_c \geq 5$, the models are confining theories and acquire a
non-perturbatively generated superpotential. By the effect of the
superpotential, there is no ground state corresponding to the vacuum
with Eq.~(\ref{eq:vacuum-2}).

In summary, there is no phenomenologically viable model with the
particle content in Table~\ref{tab:content} and superpotential in
Eq.~(\ref{eq:superpotential}), although the doublet-triplet splitting
problem is solved in a simple way.
This result motivates us to consider the case with different type of
groups such as Sp($N_c$) and SO($N_c$).

In fact, there is another interesting way of realizing massless doublet
Higgs fields in this class of models. If we impose a global SU(6)
symmetry in the superpotential where ${\cal Q} \equiv (Q, T)$ and
${\bar{\cal Q}} \equiv (\bar{Q}, \bar{T})$ transform as ${\bf 6}$ and
${\bf \bar{6}}$ and if the global symmetry is broken down to SU(4)
$\times$ SU(2) $\times$ U(1), a pair of doublet Higgs fields is ensured
to be massless since these are pseudo-Goldstone
particles~\cite{Inoue:1985cw, Cheng:1999fw, Nomura:2006pn}.
With the similar superpotential:
\begin{eqnarray}
 W = m {\rm Tr}({\cal Q} \bar{\cal Q}) 
- \frac{1}{M} {\rm Tr}[({\cal Q} \bar{\cal Q})({\cal Q} \bar{\cal Q})]
+ \cdots\ ,
\end{eqnarray}
the mechanism should work and the unwanted exotic particles can be
massive by the superpotential terms if such a vacuum exists.
Although it is an interesting possibility, we do not pursue this
direction further in this paper partly because it is incompatible with
the later discussion of supersymmetry breaking.

\subsection{Sp($N_c$) and SO($N_c$) models}

\begin{table}
\begin{center}
\begin{tabular}{ccc}
 & Sp($N_c$) & SU(5)$_{\rm GUT}$  \\ \hline \hline
 $Q$ & { $2N_c$} & {\bf 5}  \\
 ${\bar{Q}}$ & { $2N_c$} & ${\bf \bar{5}}$  \\
 $T_1$ & { $2N_c$} & {\bf 1}  \\
 $T_2$ & { $2N_c$} & {\bf 1}  \\ \hline
\hline
\end{tabular}
\end{center}
\caption{
The particle content of the Sp($N_c$) model. 
}
\label{tab:content-sp}
\end{table}

We can indeed find viable models for Sp($N_c$) and SO($N_c$).
We show the result of the analysis for these cases.
The particle content of the Sp($N_c$) models is listed in
Table~\ref{tab:content-sp} where the field $T_2$ is necessary to avoid
the Witten anomaly~\cite{Witten:1982fp}.
We assumed the same superpotential for $Q$ and $\bar{Q}$ as in
Eq.~(\ref{eq:superpotential}) with the matrix $(Q \bar{Q})$ being the
Sp($N_c$) singlet combination with $5 \times 5$ flavor indices this
time.
The analysis can go through in the similar fashion to the SU($N_c$)
case, and we find that only $N_c = 2$ case has a stable minimum with
Eq.~(\ref{eq:vacuum-2}). The massless modes of the Sp(2) model consist
of four Higgs doublets, $H_D \sim Q_D T_1 $, $\bar{H}_D \sim \bar{Q}_D
T_1$, $H_{D2} \sim Q_D T_2$ and $\bar{H}_{D2} \sim \bar{Q}_D T_2$. Two
of them $H_{D2}$ and $\bar{H}_{D2}$ can be made massive by adding a
superpotential term $W \ni (Q T_2)(\bar{Q}T_2)$ without giving a mass
for $H_D$ and $\bar{H}_D$. This is the model found in
Ref.~\cite{Kitano:2005ez}.
Similar to the SU($N_c$) case, Sp($N_c$) models with $N_c \geq 3$ do not
have a vacuum with rank$(M_{Q \bar{Q}})=2$ due to the non-perturbatively
generated superpotential~\cite{Intriligator:1995ne}.

\begin{table}
\begin{center}
\begin{tabular}{ccc}
 & SO($N_c$) & SU(5)$_{\rm GUT}$  \\ \hline \hline
 $Q$ & { $N_c$} & {\bf 5}  \\
 ${\bar{Q}}$ & { $N_c$} & ${\bf \bar{5}}$  \\
 $T$ & { $N_c$} & {\bf 1}  \\ \hline
\hline
\end{tabular}
\end{center}
\caption{
The particle content of the SO($N_c$) model. 
}
\label{tab:content-so}
\end{table}

The SO($N_c$) model can also be constructed, and turns out to be the
most interesting case. 
A detailed analysis will be presented in the next section.
The particle content is given in Table~\ref{tab:content-so} where we
have to introduce only one $T$ field in contrast to the case of
SU($N_c$) or Sp($N_c$). Again, the form of the superpotential is the
same as that in Eq.~(\ref{eq:superpotential}). With the same analysis,
we find that there are stable vacua for $ 4 \leq N_c \leq 9$ with
massless Higgs doublet fields and there are no unwanted massless fields
at low energy.
The theory is asymptotically free for $N_c \geq 6$.
Additional singlet fields under the standard model gauge group appear
for $N_c \geq 5$.

\section{SO(9) model}

We discuss in more detail the most interesting model among those in the
previous section: the SO(9) $\times$ SU(5)$_{\rm GUT}$ model.
The phenomenological aspects of the model such as gauge coupling
unification, proton decay and Yukawa interactions for matter fields will
be addressed. Many of these features are shared with the model of
Ref.~\cite{Kitano:2005ez}.

The particle content and the tree-level superpotential are defined in
the previous section in Table~\ref{tab:content-so} and in
Eq.~(\ref{eq:superpotential}). The model is an SO(9) gauge theory with
11 flavors which is in the conformal
window~\cite{Intriligator:1995id}. This fact becomes important for the
discussion of the phenomenological issues. We take a picture in which
$\Lambda \gg v$ where $\Lambda$ is the scale where the SO(9) gauge
theory flows into the fixed point. Since confinement does not happen
until the fields decouple, the actual confinement scale coincides with
the GUT scale $v$ which is set by the parameters $m$ and $M$. Therefore
we have an energy region with a CFT between $M_G \sim v$ and $\Lambda$.

The analysis of the vacuum can be done along the same line in the
SU($N_c$) case. We first take dual picture of the theory which is an
SO(6) gauge theory with 11 flavors with a
superpotential~\cite{Intriligator:1995id}:
\begin{eqnarray}
 W &=& m {\rm Tr} M_{Q \bar{Q}} 
- \frac{1}{M} {\rm Tr} (M_{Q \bar{Q}} M_{Q \bar{Q}})
+ \cdots
\nonumber \\
&+& 
\frac{1}{\hat{\Lambda}} \bar{q} M_{Q \bar{Q}} {q} 
+ \frac{1}{\hat{\Lambda}} \bar{q} M_{Q {Q}} \bar{q} 
+ \frac{1}{\hat{\Lambda}} {q} M_{\bar{Q} \bar{Q}} {q} 
\nonumber \\
&+& 
\frac{1}{\hat{\Lambda}} {H} \bar{q} {t} 
+ \frac{1}{\hat{\Lambda}} \bar{H} {q} {t} 
+ \frac{1}{\hat{\Lambda}} {S} {t} {t} \ ,
\end{eqnarray}
where $M_{Q \bar{Q}}$, $M_{Q Q}$ and $M_{\bar{Q} \bar{Q}}$ are mesons
made of $Q$ and $\bar{Q}$ which are singlet under SO(6) and ${\bf
1}+{\bf 24}$, {\bf 15} and ${\bf \overline{15}}$ under SU(5)$_{\rm
GUT}$. The mesons involving $T$ are $H \sim QT$, $\bar{H} \sim \bar{Q}T$
and $S \sim TT$, and these are ${\bf 5}$, ${\bf \bar{5}}$ and ${\bf 1}$
under SU(5)$_{\rm GUT}$. The dual quarks $q$, $\bar{q}$ and $t$
transform as $q:({\bf 6},{\bf 5})$, $\bar{q}:({\bf 6},{\bf \bar{5}})$
and $t:({\bf 6},{\bf 1})$ under SO(6) $\times$ SU(5)$_{\rm GUT}$.
Again $\hat{\Lambda}$ is introduced such that dimensionality of the
superpotential is correct. 
It is interesting that this dual picture is similar to the SO(6)
$\times$ SO(10)$_{\rm GUT}$ model proposed in Ref.~\cite{Hotta:1996qb}.
By the VEV of $M_{Q \bar{Q}}$ in Eq.~(\ref{eq:vacuum-2}), four flavors
(doublet part of $q$ and $\bar{q}$) obtain masses and the low energy
theory becomes an SO(6) theory with seven flavors (colored part of $q$
and $\bar{q}$ and $t$).  This is still an interacting theory.

When we take the dual again, the theory now comes back to the original
electric theory but the gauge group is reduced to SO(5). The
superpotential is
\begin{eqnarray}
 W &=& m {\rm Tr} M_{Q \bar{Q}} 
- \frac{1}{M} {\rm Tr} (M_{Q \bar{Q}} M_{Q \bar{Q}})
+ \cdots
\nonumber \\
&+& 
\frac{1}{\hat{\Lambda}} M_{Q \bar{Q}}^{(3 \times 3)} 
M_{q \bar{q}}^{(3 \times 3)}
 + \cdots
\nonumber \\
&-& 
\frac{1}{v^2 \hat{\Lambda}} {H_D} \bar{H}_D S^\prime
+\frac{1}{\hat{\Lambda}} {H_C} \bar{H}_C^\prime
+\frac{1}{\hat{\Lambda}} {\bar{H}_C} {H}_C^\prime
+ \frac{1}{\hat{\Lambda}} {S} S^\prime 
\nonumber \\
&-& 
\frac{1}{{\hat{\Lambda}} } 
\overline{\widetilde{Q}}_C M_{q \bar{q}}^{(3 \times 3)} \widetilde{Q}_C + \cdots
\nonumber \\
&-& 
\frac{1}{{\hat{\Lambda}} } 
\overline{\widetilde{Q}}_C H_C^\prime \widetilde{T}
- \frac{1}{{\hat{\Lambda}}} 
{\widetilde{Q}}_C \bar{H}_C^\prime \widetilde{T}
- \frac{1}{{\hat{\Lambda}}} S^\prime \widetilde{T} \widetilde{T}
\ ,
\label{eq:dual-dual}
\end{eqnarray}
where we wrote down only terms relevant for the discussion. The quarks
$\widetilde{Q}$, $\overline{\widetilde{Q}}$ and $\widetilde{T}$ will be
identified with original quarks upon integrating out the massive fields.
The doublet-triplet splitting happens in the same way as the SU($N_c$)
examples. The massless doublet is obtained with $S^\prime = 0$ which is
ensured by the $F_S = 0$ condition, and the triplets get masses by
pairing with dual mesons.
In contrast to the case of the SU(3) model in the previous section,
there is no non-perturbatively generated superpotential. Therefore, the
symmetry of the superpotential at the classical level, $T$-number
($T:+1$), is respected. This is exactly the situation discussed in the
Introduction.
This fact becomes important in the discussion of the proton decay.

By solving the $F=0$ conditions, we can find a solution in
Eqs.~(\ref{eq:vacuum-2}) and (\ref{eq:mqq-dual}) that gives a mass term
to $\widetilde{Q}_C$ and $\overline{\widetilde{Q}}_C$.
After integrating out the heavy fields, the theory ends up with an SO(5)
theory with one flavor $\widetilde{T}$ without a superpotential, and we
have massless Higgs doublet fields. This result is the same as the
classical analysis in the Higgs phase.

The decoupling of the fields immediately make the SO(5) interaction
strong and the quark $\widetilde{T}$ confines. This SO(5) one flavor
theory has known to have two branches~\cite{Intriligator:1995id}. In one
branch, a superpotential $W =
(\widetilde{\Lambda}^8/\widetilde{S})^{1/2}$ is generated with
$\widetilde{\Lambda}$ being the dynamical scale of one flavor SO(5)
theory and $\widetilde{S} \sim \widetilde{T} \widetilde{T}$. This branch
is unacceptable because there is no ground state. In the other branch,
no superpotential is generated and there is no singularity at the origin
of the meson $\widetilde{S}$ even though the gauge symmetry is enhanced
there at the classical level.
Therefore there is a stable vacuum in this branch. The low energy
spectrum is just a pair of doublet Higgs fields with a massless meson
$\widetilde{S}$ and the superpotential is $W = 0$.

Now we start to discuss the phenomenological issues. 
First, we need to check whether gauge coupling unification is maintained
in this model.
The first order answer to this question is yes. There is no exotic
massless fields in the spectrum, and the running of the gauge coupling
constants are the same as that of the minimal supersymmetric standard
model.
However, the threshold correction is also important for precise
unification, which is not obvious.
The question depends on the spectrum of the heavy fields and that
cannot be estimated without the knowledge of the K{\" a}hler potential.
However, the qualitative discussion is still possible by assuming that
the K{\" a}hler potential is not very different from the classical
one. In this case, the mass spectrum can be estimated by explicitly
calculating the mass of the components in $Q$ and $\bar{Q}$ at the
classical level. 
There are three classes of fields: fields eaten by the SO(9)/SO(5) gauge
fields, ones eaten by the SU(5)$_{\rm GUT}$/(SU(3)$_C$ $\times$
SU(2)$_L$ $\times$ U(1)$_Y$) gauge fields, and others that obtain masses
by the superpotential. The first two classes of fields have masses of
order $v = \sqrt{m M/2}$ and the last ones have $O(m)$. Therefore, in
order not to destroy gauge coupling unification due to the mass
splitting between those two, the mass parameter $m$ is required to be
around the GUT scale $M_G \sim v$ which means the scale $M$ should also
be of the order of the GUT scale.

This sounds unreasonable since we expect that the higher dimensional
operators are suppressed by the Plank scale $M_{\rm Pl}$ that is two
order of magnitude larger than the GUT scale.
Also, with GUT scale suppressed operators, we are not allowed to discuss
the high energy theory above the GUT scale, which we are doing.
However, in this model, the mechanism of suppressing $M$ is already
built in. As we discussed before, this SO(9) theory is in the conformal
window, and we expect an energy range of CFT above the GUT scale. In
this case, the meson fields have large negative anomalous dimension
which enhances the couplings in low energy. In other words, the
interaction never gets strong at high energy.
The anomalous dimension of the meson fields are calculated by using the
relation between the non-anomalous $R$-charge and the dimension of the
operator as follows~\cite{Seiberg:1994pq}:
\begin{eqnarray}
 \gamma (Q\bar{Q}) = D(Q\bar{Q}) - 2 = \frac{3}{2} R(Q\bar{Q}) - 2
= - \frac{10}{11}\ .
\end{eqnarray}
With this anomalous dimension, the coefficient $1/M$ enhances almost
quadratically with scale towards low energy. Therefore it is natural to
have a large enhancement. If we assume that the original operator is
suppressed by the Plank scale $M_{\rm Pl}$, the factor of 100
enhancement is easily realized by a small CFT range such as from
$10^{16}$~GeV to $10^{17}$~GeV.
In the same way, the coefficient $m$ enhances almost linearly with
energy scale towards low energy. Therefore, the original scale of the
model $m$ evaluated at the Planck scale was smaller than the GUT scale
by a factor of 10 or so. This is an interesting scale for the
right-handed neutrino masses in the see-saw model~\cite{seesaw}.

The gauge coupling of SU(5)$_{\rm GUT}$ above the GUT scale can be in
the perturbative region all the way up to the Planck scale even
accounting for the large anomalous dimension of $Q$ and
$\bar{Q}$~\cite{Novikov:1983uc}.

The Yukawa coupling constants between matter and the Higgs fields
originate from higher dimensional operators since the Higgs fields are
identified with the meson fields. The gauge invariant terms:
\begin{eqnarray}
 W_{\rm Yukawa} = \frac{f_u}{M_Y} ({\bf 10}) ({\bf 10}) (Q T) 
+ \frac{f_d}{M_Y} ({\bf 10}) ({\bf \bar{5}}) (\bar{Q} T)\ ,
\label{eq:Yukawa}
\end{eqnarray}
become the Yukawa interactions at low energy. The matter fields are
represented by (${\bf 10}$) and (${\bf \bar{5}}$). The low energy Yukawa
coupling is roughly $y_u \sim f_u M_G / M_Y$ and $y_d \sim f_d M_G /
M_Y$, where $M_G$ is the GUT scale.
The Yukawa coupling constant for the top quark is necessary to be
$O(1)$, which again requires that the scale $M_Y$ to be the GUT
scale. This is not a problem for the same reason as before. These
operators are almost marginal operators and thus the coefficient is
enhanced linearly in low energy by the large anomalous dimensions.

After integrating out the massive fields with the Yukawa interactions in
Eq.~(\ref{eq:Yukawa}), the final low energy effective superpotential is:
\begin{eqnarray}
 W = W_{\rm MSSM}
+ \frac{y_u y_d }{m}
\frac{\widetilde{S}}{M_G^2} ( 
  \hat{Q} \hat{Q} \hat{Q} \hat{L} 
+ \hat{U} \hat{U} \hat{D} \hat{E} 
+ \hat{Q} \hat{Q} \hat{U} \hat{D} 
+ \hat{U} \hat{E} \hat{Q} \hat{L} 
) \ ,
\label{eq:eff-sup}
\end{eqnarray}
with $\hat{Q}$, $\hat{U}$, $\hat{D}$ being the quark superfields and
$\hat{L}$, $\hat{E}$ are the lepton superfields.  In addition to the
usual Yukawa interactions $W_{\rm MSSM}$, the baryon-number-violating
terms (the first two terms in the parenthesis) appeared.  However, as
long as $\widetilde{S}$ is stabilized near the origin, these terms do
not cause rapid proton decay.
The particle content of the low energy effective theory is just that of
the minimal supersymmetric standard model and a singlet field
$\widetilde{S}$ which only couples to the higher dimensional operators.
The value of $\widetilde{S}$ depends on the shape of the K{\"
a}hler potential and how supersymmetry is broken.
We will discuss these in the next section.

We briefly mention the case with $N_c = 6,7$ and 8, where the vacuum
with Eq.~(\ref{eq:vacuum-2}) exists and the doublet-triplet splitting
happens for these cases. All of the models are in the conformal window.
For $N_c = 6$, the low energy effective theory has an extra U(1) gauge
symmetry with two charged (but the standard model singlet) fields $t^+$
and $t^-$. Since this theory is at the edge of the conformal window
(barely asymptotically free), we do not expect the large enhancement of
the Yukawa coupling constants. For $N_c = 7$, the low energy theory
again has an extra U(1) factor. There are three standard model singlet
fields $S$, $t^+$ and $t^-$, where $t^+$ and $t^-$ are monopoles, with
superpotential $W = f(S) S t^+ t^-$. The unknown function $f(S)$ is
non-vanishing at the origin. For $N_c = 8$, there is no extra gauge
symmetry and there are two branches as in the $N_c=9$ case. In one
branch, stable vacuum exists with a superpotential term $W = f(x) S t t$
where $S$ and $t$ are the standard model singlet fields. Again, $f(x)$
($x = S^2 t^2$) is an unknown function but non-zero at the origin.

\section{{\boldmath $\mu$}-term driven supersymmetry breaking}

In the previous section, we have seen that the SO(9) model is quite
successful in obtaining massless Higgs fields in the low energy
spectrum. However, to be phenomenologically completely viable, the Higgs
fields have to have a mass term of $O(100~{\rm GeV})$, the $\mu$-term,
otherwise Higgsinos become massless which is excluded by experiment. As
we see below, it is possible to obtain a $\mu$-term by giving a small
mass term for $T$, but it causes a dramatic effect in the dynamical
system: dynamical supersymmetry breaking~\cite{Witten:1981nf,
Affleck:1983rr}.

The $\mu$-term can be obtained by adding a superpotential term:
\begin{eqnarray}
 W \ni \hat{\mu} T T\ ,
\label{eq:mu-hat}
\end{eqnarray}
which becomes $\hat{\mu} S$ in Eq.~(\ref{eq:dual-dual}). With this term,
the $F_S = 0$ condition leads to
\begin{eqnarray}
 S^\prime = - \hat{\mu} \hat{\Lambda}\ ,
\label{eq:s-prime}
\end{eqnarray}
and it induces a mass term for the Higgs doublets:
\begin{eqnarray}
 W \ni \frac{\hat{\mu}}{v^2} H_D \bar{H}_D \ .
\end{eqnarray}
In terms of the canonically normalized fields $\hat{H}_D$ and
$\hat{\bar{H}}_D$, this is nothing but the $\mu$-term, $\mu \hat{H}_D
\hat{\bar{H}}_D$ with $\mu \sim \hat{\mu} $.
It is obvious that the mesons made of $T$ become massive once we
introduce the mass term for the quark $T$.
This is also easy to understand in the classical level analysis. Since
the Higgs fields are simply the components in $T$, the term in
Eq.~(\ref{eq:mu-hat}) is the $\mu$-term.

The term in Eq.~(\ref{eq:mu-hat}) also gives a potential term for
$\widetilde{S}$. With Eq.~(\ref{eq:s-prime}) and the superpotential in
Eq.~(\ref{eq:dual-dual}), $\widetilde{T}$ obtains a term in low energy:
\begin{eqnarray}
\hat{\mu} \widetilde{T}\widetilde{T}\ ,
\end{eqnarray}
and after confinement of the SO(5) gauge theory, it becomes a linear
superpotential for the meson $\widetilde{S} \sim \widetilde{T}
\widetilde{T}$:
\begin{eqnarray}
 W \ni \hat{\mu} \widetilde{S}\ .
\end{eqnarray}
If we ignore the higher dimensional operators in Eq.~(\ref{eq:eff-sup}),
there is no solution for $F_{\tilde{S}} = 0$.

Interestingly, this does not mean that the vacuum is destabilized or
quarks and leptons must condense by the presence of the small
$\mu$-term.
First, we start the discussion by ignoring the higher dimensional
operators.
In this case, it was shown by Intriligator, Seiberg and
Shih~\cite{Intriligator:2006dd} that the vacuum is meta-stable and
supersymmetry is broken there. The argument is pretty easy. Whether the
vacuum is stable or not depends on the shape of the K{\" a}hler
potential for $\widetilde{S}$, but since we know that for large
$\widetilde{S}$, where the classical analysis is valid, the potential
grows by the mass term. Therefore, there must be a local minimum
somewhere.
The true supersymmetric vacua exists in the other branch where
$\tilde{S}$ is stabilized far away from the origin $M_G (M_G /
\hat{\mu})^{2/3}$ (this is meaningless because it is much larger than
the Planck scale) and also at different vacua from that in
Eq.~(\ref{eq:vacuum-2}). Since the true vacua exist far away or energy
difference between the true vacuum and the meta-stable one is much
smaller than the height of the potential barrier of $O(M_G^4)$, we
expect that the life-time of this meta-stable vacuum is long enough for
us~\cite{Coleman:1977py}.

Once we include the higher dimensional operators, another supersymmetric
vacuum appears where quarks and leptons acquire VEVs. However, if we
assume that the scalar components of quarks and leptons obtain positive
supersymmetry breaking mass-squared terms, the vacuum is again
meta-stable and its life-time is very long since the peak of the
potential barrier is located far from the origin $Q \sim L \sim (\mu
M_G)^{1/2}$ compared to the height of the potential $V^{1/4} \sim
O(\mu^3 M_G)^{1/4}$~\cite{Duncan:1992ai}.
Therefore, we conclude that there is a supersymmetry breaking
meta-stable vacuum.

The size of supersymmetry breaking is $F_{\hat{S}} \sim O(\mu M_G)$ with
a canonically normalized field $\hat{S} \sim M_G \widetilde{S}$.
For the Higgs fields, terms in the K{\" a}hler potential such as
$\hat{S}^\dagger \hat{S} H^\dagger H / \tilde{\Lambda}^2$ are expected
to be generated by the non-perturbative effect and supersymmetry
breaking can be mediated directly (in the sense of gravity
mediation). In this case, the soft scalar masses for the Higgs fields
are obtained with a similar size to the $\mu$-term.
If the term $\hat{S} H^\dagger H / \tilde{\Lambda}+ {\rm h.c.}$ is
generated, which should be possible since there is no unbroken symmetry
to protect the term, the trilinear $A$ and the bilinear $B$-term is also
non-vanishing and of the same order with the $\mu$-term. If the matter
fields in the third generation couple to the Higgs fields strongly, the
soft scalar masses for those fields can also be obtained directly.

Gauge mediation~\cite{Dine:1994vc} also happens if ${\hat{S}}$ is
stabilized away from the origin, where the colored-Higgs fields play a
role of the messenger field\footnote{A similar structure of the model
can be found in Ref.~\cite{Izawa:2005yf}.}. By integrating out
${\widetilde{Q}}_C$ $\overline{\widetilde{Q}}_C$ in
Eq.~(\ref{eq:dual-dual}), we obtain mass terms for the colored-Higgs
fields:
\begin{eqnarray}
 W \simeq 
\hat{S} \hat{H}_C^\prime \hat{\bar{H}}_C^\prime
+ M_C {\hat{H}_C} \hat{\bar{H}}_C^\prime
+ M_C \hat{\bar{H}}_C \hat{H}_C^\prime
\ ,
\end{eqnarray}
where $M_C$ is the colored-Higgs mass of order $M_G$ and meson fields
are canonically normalized. 
Unfortunately, with this structure of superpotential, the leading order
contribution to the gaugino masses of $O( F_{\hat{S}} / \langle \hat{S}
\rangle)$ cancels out~\cite{Izawa:1997gs}\footnote{I thank Y.~Nomura for
discussion on this point.}, and moreover there is no contribution to the
SU(2)$_L$ gauginos.

In order to obtain gaugino masses, there must be a gauge-mediation
effect since the supersymmetry breaking scale is too low ($O(\mu M_G)$)
for gravity mediation.
A simple example for gauge mediation is to assume an interaction term:
\begin{eqnarray}
 W_{\rm messenger} = \frac{1}{M_X} T^2 \Phi \bar{\Phi} 
\ ,
\end{eqnarray}
where $\Phi$ and $\bar{\Phi}$ transform under the SU(5)$_{\rm GUT}$ such
as ${\bf 5}$ and ${\bf \bar{5}}$ and singlet under SO(9).
By the enhancement of the $1/M_X$ suppressed term due to large anomalous
dimension of $T^2$, this term effectively becomes
\begin{eqnarray}
 W_{\rm messenger} \to \lambda \hat{S} \Phi \bar{\Phi} 
\end{eqnarray}
with $\lambda \sim O(0.1-1)$ even if $M_X$ is $O(M_{\rm Pl})$.  
With this superpotential, another supersymmetric true vacuum with $\Phi
= \bar{\Phi} \neq 0$ appears. However, assuming that the $\hat{S}$ field
is stabilized away from the origin, which is reasonable since we expect
the presence of a linear term in the K{\" a}hler metric, the vacuum with
$\Phi = \bar{\Phi} = 0$ is meta-stable\footnote{Since the linear term in
the K{\" a}hler metric has non-vanishing $R$-charge, it is suppressed
when the explicit $R$-symmetry breaking by the superpotential is small,
i.e., $m \ll \widetilde{\Lambda}$. Even in this case, the supergravity
effect shifts the vacuum to $\langle \hat{S} \rangle \sim
\widetilde{\Lambda}^2/ M_{\rm Pl} \sim 10^{14}$~GeV, which is
numerically consistent with the phenomenological requirements. See
\cite{Kitano:2006wz} for detailed discussion.}.
At the meta-stable vacuum the gaugino masses are generated to be
\begin{eqnarray}
m_{1/2} = \frac{\alpha}{4 \pi} 
\frac{F_{\hat{S}}}{\langle \hat{S} \rangle} \ .
\label{eq:gauge-med}
\end{eqnarray}
In order for the gaugino masses to be similar to $\mu$ in size, somewhat
small value of $ \langle \hat{S} \rangle \sim 10^{14-15}$~GeV is
necessary. That is consistent with the suppression of the coefficient of
proton-decay operators in Eq.~(\ref{eq:eff-sup}).

It might be possible and would be great if all the gaugino masses are
obtained by (really) direct gauge mediation~\cite{Poppitz:1996fw}
without having the messenger particles above by extending the gauge
group and/or matter content.
But in any case, the pattern of the supersymmetry breaking parameters in
this scenario is essentially of gauge-mediation type except for the
Higgs sector, since the Higgs fields can feel the supersymmetry breaking
directly. The soft scalar masses $m_{H_u}^2$, $m_{H_d}^2$, $\mu$, $B$
and $A$-terms can be taken as free parameters at the GUT
scale. (Probably $m_{H_u}^2 \simeq m_{H_d}^2$ because of the parity
symmetry $Q \leftrightarrow \bar{Q}$ which is only broken by the Yukawa
interactions.) The soft masses for the third generation fields may also
be modified depending on the size of the Yukawa coupling constants. This
prediction should be testable at future colliders.
The source of flavor and CP violation in this model is only in the
Yukawa coupling constants, which is the desired situation taking into
account the stringent constraints on the supersymmetry breaking
parameters.
The gravity-mediation effect gives an $O(1\%)$ correction to the
parameters. This is interesting for the detection of the flavor and CP
violating processes\footnote{One should take into account the possible
conformal sequestering effect~\cite{Luty:2001jh} to the gravity-mediated
contribution.  }.

The lightest supersymmetric particle is the gravitino. The gravitino
mass is estimated to be
\begin{eqnarray}
 m_{3/2} = \frac{F_{\hat{S}}}{\sqrt{3} M_{\rm Pl}}
\sim \mu \left(
\frac{M_G}{M_{\rm Pl}}
\right)
\sim O(1)~{\rm GeV}
\ .
\label{eq:gravitino}
\end{eqnarray}
This mass range is interesting for cosmology~\cite{Feng:2003xh} and also
for collider experiments~\cite{Buchmuller:2004rq}.

For other choices of $N_c$, the situation is different. For $N_c = 6$,
the addition of the $\mu$-term just gives a mass term for $t^+$ and
$t^-$ which does not cause supersymmetry breaking. For $N_c = 7$ and 8,
the term $\mu S$ appears but the supersymmetry unbroken vacuum exists
where $t^{\pm}$ (or $t$ for $N_c = 8$) fields acquire non-vanishing
VEV. However, it is possible that the $S$ is stabilized far from the
origin where $t^{\pm}$ or $t$ is heavy, and supersymmetry is broken
there.

%\section{Gravitational supersymmetry breaking}
\section{Cosmological constant driven supersymmetry breaking}

Because of the large anomalous dimension, the size of $\mu$-parameter is
originally smaller than $O(100~{\rm GeV})$ by a factor of 10 or so. This
is about the size of the gravitino mass. 
Therefore it is possible that the origin of the $\mu$-term can actually
be the cosmological constant by the Giudice-Masiero
mechanism~\cite{Giudice:1988yz}. 
Assuming a presence of a $T^2$ ($\sim \widetilde{\Lambda} \hat{S}$) term
in the K{\" a}hler potential, this effectively becomes the
$\hat{\mu}$-term in the superpotential.
It is equivalent to study a model with
\begin{eqnarray}
 K = \hat{S}^\dagger \hat{S} + a \widetilde{\Lambda} ( \hat{S} + {\rm h.c.} ) 
- \frac{(\hat{S}^\dagger \hat{S})^2}{\widetilde{\Lambda}^2} 
+ \cdots \ \ \ {\rm and}\ \ \ 
 W = c\ ,
\end{eqnarray}
where $c$ is a constant term which represents the negative cosmological
constant term, $V_{\rm AdS} \simeq -3 |W|^2$, of order $c^2 \sim m_{3/2}^2$ in the unit
of $M_{\rm Pl} = 1$.
This term is always necessary to cancel the positive vacuum energy from
the supersymmetry breaking, $V_F \simeq |F|^2$.
The parameters have a hierarchical structure: $a \gg 1$,
$\widetilde{\Lambda} \ll 1$ and $c \ll 1$, where $a$ represents the
enhancement of the coupling through the large anomalous dimension.

By the K{\" a}hler transformation, $K \to K - x - x^\dagger$ and $W \to
W e^x$ with $x$ being a chiral superfield, this is identical to the
system:
\begin{eqnarray}
 K = \hat{S}^\dagger \hat{S}
- \frac{(\hat{S}^\dagger \hat{S})^2}{\widetilde{\Lambda}^2} 
+ \cdots \ \ \ {\rm and}\ \ \ 
 W = c e^{a \widetilde{\Lambda} \hat{S}} \ .
\end{eqnarray}
By expanding the superpotential, we obtain a $\hat{\mu}$-term of order
$a c \simeq a m_{3/2}$.
Within the range $|\hat{S}| \lesssim \widetilde{\Lambda}$, where the
effective theory makes sense, the minimum of the potential exists near
the origin $\hat{S} \simeq \widetilde{\Lambda}/(4a)$, and the
cosmological constant can be cancelled when $a \simeq
\sqrt{3}/\widetilde{\Lambda} \sim O(100)$.
Supersymmetry is broken at the minimum with $F_{\hat{S}} \simeq ac
\widetilde{\Lambda} \simeq \mu \widetilde{\Lambda}$.
The value of $\hat{S}$ at the minimum is $O(10^{14}~{\rm GeV})$ which is
consistent with the required value for gauge mediation in
Eq.~(\ref{eq:gauge-med}) and also the suppression of the coefficient of
the dimension-five proton-decay operators in Eq.~(\ref{eq:eff-sup}).

In the conventional scenario of supersymmetry breaking, some mass
scales, such as a dynamical scale, determine the size of $|F|^2$ and the
net cosmological constant is cancelled by an independent negative
contribution from the $c$-term. However, in this scenario, the $c$-term
drives supersymmetry breaking, and therefore these are related. In
particular, it is interesting that supersymmetry is recovered in the
flat limit ($c \to 0$), resulting in the supersymmetric flat space
rather than the supersymmetry broken de Sitter space.

\section{Discussion and Conclusions}

From the consideration of the mystery of the Higgs particle, we arrived
at a rather unified picture. At every stage of the phase transitions,
GUT breaking, supersymmetry breaking and electroweak symmetry breaking,
the Higgs field may be playing a crucial role. We have succeeded to
construct a realistic GUT model with dynamical symmetry breaking, and
found that, in the SO(9) model, the inclusion of the $\mu$-term for the
Higgs fields triggers supersymmetry breaking in a meta-stable vacua by
the same dynamics.

We discuss possible generalizations of the model here.
Although we discussed the GUT breaking and supersymmetry breaking in a
unified picture, we can separately discuss the following mechanisms:
\begin{itemize}
 \item Doublet-triplet splitting through dynamical GUT breaking,
 \item $\mu$-term driven supersymmetry breaking.
\end{itemize}

Dynamical GUT breaking without supersymmetry breaking is possible. The
$\mu$-term can be obtained from separate SUSY breaking sector by, e.g.,
the Giudice-Masiero mechanism.
The SO(10)$_{\rm GUT}$ extension of the model should be straightforward
and is interesting for the discussion of the neutrino masses.
Considering different types of particle content and assumption on the
superpotential is also worth investigating.

It is possible to obtain massless colored Higgs fields instead of the
doublet fields.
In this case, the doublet-triplet splitting can be done by introducing a
pair of elementary Higgs fields with the coupling to the meson
operators~\cite{Yanagida:1994vq,Hotta:1995ih,Hotta:1996qb,Izawa:1997br}.
This is possible in the rank($M_{Q \bar{Q}}$)=3 vacuum with SO($N_c$)
with $6 \leq N_c \leq 11$.

$\mu$-term driven supersymmetry breaking can be discussed without
GUT. For example, in the SO(9) model we can gauge only SU(2)$_L$
$\times$ U(1)$_Y$ subgroup of SU(5)$_{\rm GUT}$. Then the dynamical
scale can be lowered (or even raised) as long as the gauge couplings of
the standard model gauge group maintains the perturbativity.
Adding a $\mu$-term breaks supersymmetry in the same way but the scale
can be much lower (or higher).
Gauge mediation through the colored-Higgs fields might be able to be
generalized as a realistic direct gauge-mediation model. We leave those
questions for future studies.
In any case, the prediction to the low energy spectrum is a modification
of the gauge-mediation type in the Higgs sector.
In the SO(9) model we presented, there is a cosmological problem
associated with the modulus $\hat{S}$~\cite{Banks:1993en}. The mass of
$\hat{S}$ is of the same order of $\mu$, i.e., $O(100~{\rm GeV})$
independent of the dynamical scale. A realistic cosmological scenario
needs to be considered.

We see that two mechanisms non-trivially fit into a picture: dynamical
GUT and supersymmetry breaking. Although it is not likely that we can
directly probe the GUT theory by experiments, the spectrum of the
supersymmetric particles in low energy gives us a hint for high energy
theories. In this model, the direct connection between the Higgs fields
and the supersymmetry breaking sector provides a characteristic feature
in the low energy spectrum.

\section*{Acknowledgments}

I thank Roni Harnik for reading the manuscript.
This work was supported by the U.S. Department of Energy under contract
number DE-AC02-76SF00515.

\end{document}